\documentclass[a4paper,11pt]{article}
\pdfoutput=1 % if your are submitting a pdflatex (i.e. if you have
\usepackage{jheppub} % for details on the use of the package, please
\usepackage[T1]{fontenc} % if needed

\usepackage[utf8x]{inputenc}

\usepackage{pgf,tikz}
\usetikzlibrary{arrows}

\usepackage{graphicx} % Required for including images
\graphicspath{{Images/}} % Location of the graphics files

\usepackage{todonotes}
\title{Classical Holographic Codes}

\author[a]{Enrico M. Brehm}
\author[a,b,c]{and Benedikt Richter}
\affiliation[a]{Arnold Sommerfeld Center for Theoretical Physics, Fakul\"at f\"ur Physik, Ludwig-Maximilians-Universit\"at M\"unchen, Germany}
\affiliation[b]{Instituto de Telecomunica\c{c}\~oes, Physics of Information and Quantum Technologies Group, Portugal}
\affiliation[c]{Instituto Superior T\'{e}cnico, Universidade de Lisboa, Portugal}

\emailAdd{e.brehm@physik.uni-muenchen.de}
\emailAdd{benedikt.richter@physik.uni-muenchen.de}
\preprint{LMU-ASC 44/16}

\abstract{In this work, we introduce classical holographic codes. These can be understood as concatenated probabilistic codes and can be represented as networks uniformly covering hyperbolic space. In particular, classical holographic codes can be interpreted as maps from bulk degrees of freedom to boundary degrees of freedom. Interestingly, they are shown to exhibit features similar to those expected from the AdS/CFT correspondence. Among these are a version of the Ryu-Takayanagi formula and intriguing properties regarding bulk reconstruction and boundary representations of bulk operations. We discuss the relation of our findings with expectations from AdS/CFT and, in particular, with recent results from quantum error correction.}

\begin{document} 
\maketitle
\flushbottom

\section{Introduction}

The holographic principle is the statement that a gravitational theory describing a region of space (the bulk) is equivalent to a (nongravitational) theory confined to the boundary of that region \cite{'tHooft:1993gx,Susskind:1994vu}. That is, intrinsically nongeometric features can be equivalently described geometrically. An explicit and very well understood example is the AdS/CFT correspondence \cite{Maldacena:1997re}. It relates (quantum) gravity on $(d+1)$-dimensional asymptotically Anti-de Sitter (AdS) space to a $d$-dimensional conformal field theory (CFT) on the boundary. One remarkable aspect of this duality is the interplay of geometry and entanglement that is most evident in the proposal by Ryu and Takayanagi that entanglement entropy in the CFT is equivalently given by the area of a minimal surface in the AdS geometry \cite{Ryu:2006bv, Ryu:2006ef}. This is known as the Ryu-Takayanagi (RT) formula.

Since then, many more connections between geometry and entanglement have been proposed \cite{Swingle:2009bg, VanRaamsdonk:2010pw, Maldacena:2013xja,  Hayden:2011ag,
   Lashkari:2013koa, Brown:2015bva,  Freedman2016zud}. Also, more generally, concepts of quantum information theory were fruitfully applied to gravity  and, in particular, to black holes \cite{Hayden:2007cs, Braunstein:2009my, Almheiri:2012rt, Harlow:2013tf, Dvali:2016zqx, Dong:2016fnf, Dvali:2016lnb}. Recently, tensor networks  -- a tool originally from condensed matter physics to efficiently represent quantum many-body states, especially their entanglement structure \cite{Vidal:2007hda} -- were employed to describe holography \cite{Qi:2013caa} and, in particular, AdS/CFT \cite{Swingle:2009bg, Swingle:2012wq}. Furthermore, similarities between the properties of bulk operator reconstruction in AdS/CFT and properties of certain quantum error-correcting codes (QECC) were reported in \cite{Almheiri:2014lwa}. There, it is argued that operator reconstruction properties of AdS/CFT are captured by the fact that bulk logical operations can be described by multiple operations on the boundary. Implementing these ideas, an interesting family of toy models for holography was proposed in \cite{Pastawski:2015qua}. There, the authors combine tensor networks and quantum error-correcting codes. AdS space is tiled with perfect tensors that build up a holographic code and establish an isometric tensor from the bulk to the boundary. These holographic quantum error-correcting codes reproduce some of the key features of the AdS/CFT correspondence, as \textit{e.g.} the RT formula and remarkable bulk reconstruction properties. Later, it was pointed out that a version of the Ryu-Takayanagi formula holds quite generically in quantum error-correcting codes \cite{Harlow:2016vwg}. Furthermore, networks of random tensors \cite{Hayden:2016cfa, Nezami:2016zni} and almost-perfect tensors \cite{Bhattacharyya:2016hbx} were considered.  Also, issues like sub-AdS locality \cite{Yang:2015uoa} and the relation to gauge invariance \cite{Mintun:2015qda} were addressed.  All these constructions are intrinsically quantum and focus on the structure of entanglement.

In this work, we pose the question how far one can get without quantum correlations, like entanglement. Or to put it differently, which features can be reproduced in classical codes that are defined on similar networks? Interestingly, we are able to produce features similar to many of those mentioned above. Thereby, one of the goals is to emphasize and clarify the importance of the structure of (classical) correlations in holographic models. However, even so that we obtain important properties that are expected to hold in AdS/CFT, we do not intend to develop a purely classical model for the AdS/CFT correspondence.

First, motivated by the qutrit example provided in \cite{Almheiri:2014lwa}, we consider a classical encoding for trits, where one logical (bulk) trit is probabilistically encoded in three  physical (boundary) trits. This code has the properties that a ``version of the RT formula'' for the mutual information holds, the bulk trit can be reconstructed from any two of the boundary trits and logical operations on the bulk trit can be represented by operations on any two of the boundary trits, {\it i.e.}, there is a notion of subregion duality. Therefore,  key features of the AdS/CFT correspondence are captured qualitatively\footnote{By ``qualitatively'' we mean that, for example, we cannot represent general quantum operators, as the system is classical. However, we can implement all classical logical operations on the bulk trit by acting on a subset of the boundary trits.} by this example.

Motivated by this example, we then construct a classical code on a network defined by a uniform tiling of hyperbolic space, inspired by the holographic quantum error-correcting codes of \cite{Pastawski:2015qua}. We reproduce many of the features of this quantum code but phrased in a classical language. Although the code is classical, it is not deterministic. We choose probabilistic mappings at the vertices of the network and therefore the full mapping from bits in the bulk to bits on the boundary is probabilistic, too. The code produces entropy and classical correlations, where   we focus on the latter. For example, we compare the result for the classical mutual information -- a measure of correlations -- of a finite interval on the boundary with the result for the quantum mutual information. We find that in our classical examples a version of the RT formula holds. That is, the mutual information of an interval on the boundary and its complement is directly proportional to the length of the corresponding minimal surface in the bulk. On the one hand, it might be suspected that the mutual information scales as the area of the minimal surface, since the entanglement entropy measures both, classical and quantum correlations, and it scales with the area of this minimal surface.  On the other hand, it is, a priori, not clear, as we do not require any ``quantumness" at all to produce the result.\footnote{One might argue that the randomness required to generate a probabilistic mapping is reminiscent of quantum superpositions. However, in the case we study, there is no need for any quantum correlations and the randomness in any probability distribution could, in principle, be interpreted as arising from some quantum superposition. } This points to the fact that the structure of all correlations, classical and quantum, is encoded in the underlying geometric structure.

We also investigate the reconstruction of bulk bits  from the knowledge of subsets of the boundary bits and the representation of bulk logical operations on the boundary. Bit flips on a single bulk bit correspond to nonlocal operations on the boundary. Furthermore, there exists a notion of subregion duality. Therefore, we find a remarkable similarity to the results from quantum codes modeling holography. 

The rest of this work is organized as follows. In section \ref{Holographic quantum error-correcting codes}, we briefly review the properties of the qutrit error-correcting code introduced in \cite{Cleve:1999qg} and the holographic pentagon code of \cite{Pastawski:2015qua}. Next, in section \ref{Classical holographic codes}, we introduce {\it classical holographic codes}. We begin by analyzing a probabilistic trit code that resembles many AdS/CFT-like features, in section \ref{simpleexcl}. Subsequently, in section \ref{fullnetwork}, we study a network, where each vertex is interpreted as a probabilistic mapping. In particular, in section \ref{features}, we prove a version of the RT formula for the mutual information, the possibility of bulk reconstruction from regions on the boundary, the representation of bulk operations on the boundary, and subregion duality. In addition, we discuss the secret sharing property of these codes. In section \ref{Interpretation}, we give a physical interpretation of the radial direction in the bulk as a coarse graining parameter that interpolates between the macroscopic description in the center of AdS and the microscopic description on the boundary. Finally, in section \ref{Conclusions}, we give the conclusions of this work.

\section{Holographic quantum error-correcting codes}\label{Holographic quantum error-correcting codes}

\subsection{Qutrit example}\label{simpleexqu}

In this section, we briefly review a very simple toy model for the AdS/CFT correspondence that is based on quantum error correction. It is formulated as a qutrit\footnote{A qutrit is very similar to a qubit. However, there is one additional base vector spanning its Hilbert space. Therefore, the qutrit state is described by $|\psi\rangle=\sum_{i=0}^2 c_i | i \rangle$.} code that encodes one logical qutrit into three physical ones such that the logical qutrit can be reconstructed even if one of the physical ones is lost. The key idea is to identify the bulk degrees of freedom with logical qutrits and the boundary degrees of freedom with the physical qutrits \cite{Almheiri:2014lwa}. The logical qutrit $|\tilde{\psi}\rangle$ is encoded as 
\begin{equation}\label{qutritencoding}
\begin{split}
|\tilde{0}\rangle=&\frac{1}{\sqrt{3}}\left(|000\rangle+|111\rangle+|222\rangle\right)\,,\\
|\tilde{1}\rangle=&\frac{1}{\sqrt{3}}\left(|012\rangle+|120\rangle+|201\rangle\right)\,,\\
|\tilde{2}\rangle=&\frac{1}{\sqrt{3}}\left(|021\rangle+|102\rangle+|210\rangle\right)\,,
\end{split}
\end{equation}
where we indicated the logical qutrit by a tilde to distinguish it from the physical ones \cite{Cleve:1999qg}. That is, the logical qutrit is encoded in a subspace of the larger Hilbert space of three qutrits, where the code subspace is spanned by the GHZ-type states (\ref{qutritencoding}),
\begin{equation}
|\tilde{\psi}\rangle=\sum_{i=0}^2 c_i | \tilde{i} \rangle\,.
\end{equation} 
In consequence, none of the physical qutrits can carry any information about the encoded state, as its reduced density matrix is maximally mixed. However, interestingly, from any two physical qutrits we denote them by $A$, $B$ and $C$, the logical one can be reconstructed. That is because of the existence of operators $U_{IJ}$, where $I,J=A,B,C$, acting nontrivially only on two of the physical qutrits such that
\begin{figure}[ht]
\centering
\includegraphics{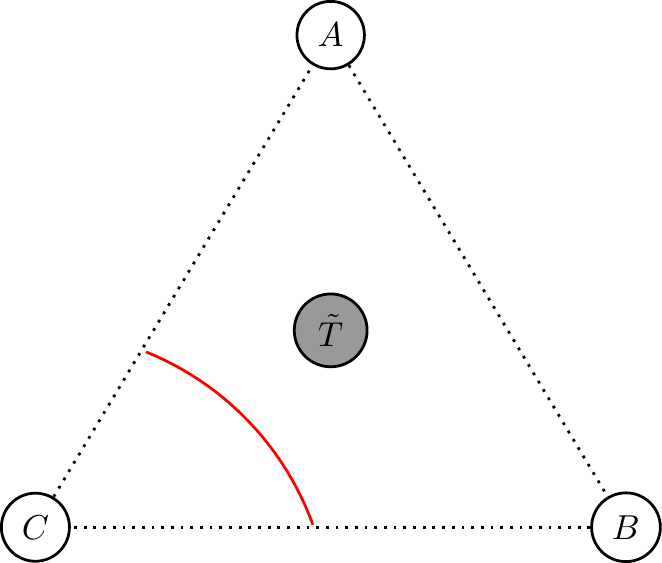}
\caption{Simplistic toy model for the AdS/CFT correspondence. One logical qutrit $\tilde{T}$ (representing the bulk degrees of freedom) is encoded in three physical qutrits $A$, $B$ and $C$ (representing the boundary degrees of freedom). The red line sketches a minimal surface in the bulk. The logical qutrit can be reconstructed from any two of the boundary qutrits, while only one of these contains no information about it. Furthermore, logical operations on $\tilde{T}$ can also be performed by acting on only two of the physical qutrits. These features are also captured in a classical version of this code we introduce in section \ref{simpleexcl}.} \label{qutrit}
\end{figure}
\begin{equation}
U_{IJ} |\tilde{i}\rangle=|i\rangle_I\otimes|\chi\rangle_{JK}\,, \hspace{15mm} |\chi\rangle=\frac{1}{\sqrt{3}}\left(|00\rangle+|11\rangle+|22\rangle\right)\,.
\end{equation}
Therefore it is clear that access to any two qutrits out of the three ($I, J, K \in \{A,B,C\}$) suffices to learn about the logical qutrit. One simply acts on these two physical qutrits with the operator $U_{IJ}$ and obtains qutrit $I$ in the state $|i\rangle$ of the logical qutrit. From this it follows that the action of a logical operator $\tilde{O}$, acting as $\tilde{O}|\tilde{i}\rangle=\sum_j \tilde{O}_{ji}|\tilde{j}\rangle$, can be achieved by the action of a corresponding operator $O_{IJ}$ acting nontrivially on any two physical qutrits. It is of the form
\begin{equation}
O_{IJ}=U_{IJ}^\dagger O_I U_{IJ}\,,
\end{equation}
where $O_I$ denotes an operator acting solely on qutrit $I$ such that $O_{IJ}| \tilde{i} \rangle=\sum_j \tilde{O}_{ji}|\tilde{j}\rangle$. That is, any logical operation $\tilde{O}$ on the logical qutrit can be performed by acting with the corresponding $O_{IJ}$ on any two physical qutrits. As it was pointed out in \cite{Almheiri:2014lwa}, this models ``subregion duality'' in AdS/CFT. Furthermore, this simple toy model obeys a version of the RT formula \cite{Harlow:2016vwg}, as we demonstrate next.

As it is clear from above, an arbitrary (mixed) state $\tilde{\rho}$ on the code subspace can be written as
\begin{equation}\label{statequtrit}
\tilde{\rho}= U_{AB} \Big(\rho_A\otimes |\chi\rangle\langle \chi|_{BC}\Big)  U_{AB}^\dagger\,.
\end{equation}
Interpreting the physical qutrits $A$, $B$ and $C$ as boundary degrees of freedom, we can calculate the entanglement entropy between  regions (here: points) in the boundary, see figure \ref{qutrit}. From (\ref{statequtrit}), one easily obtains  the entanglement entropies
\begin{align}
S(\tilde{\rho}_C)=& \log(3)\,,\nonumber\\
S(\tilde{\rho}_{AB})=& \log(3)+S(\tilde{\rho})\,,
\end{align}
where $\tilde{\rho}_C$ and $\tilde{\rho}_{AB}$ are the reduced density matrices of qutrits $C$ and $AB$, respectively. That fulfills the RT-formula with area operator $\log(3)$ \cite{Harlow:2016vwg}. Closely related to entanglement entropy is the mutual information that is, in the present case, given by
\begin{equation}\label{quantummutinfo}
I_{qu}(C,AB)= S(C)+S(AB)-S(C, AB)=2\log(3)\,.
\end{equation}
The mutual information, however, does not capture contributions from the bulk entropy in this model.\footnote{This does not necessarily hold for more elaborate models, as bulk matter can backreact and, in principle, it can modify the geometry.}   Therefore, restricting the states of the boundary qutrits to the class of pure states it is evident that the RT formula can be stated in terms of the mutual information $I_{qu}(A,A^c)$. In this form the RT formula states that the mutual information between a boundary region $A$ and its complement $A^c$ is given by twice the area of the minimal surface in the bulk.

\subsection{Holographic pentagon code}\label{Pentagon_code}

\begin{figure}[]
\centering
\includegraphics[width=8cm]{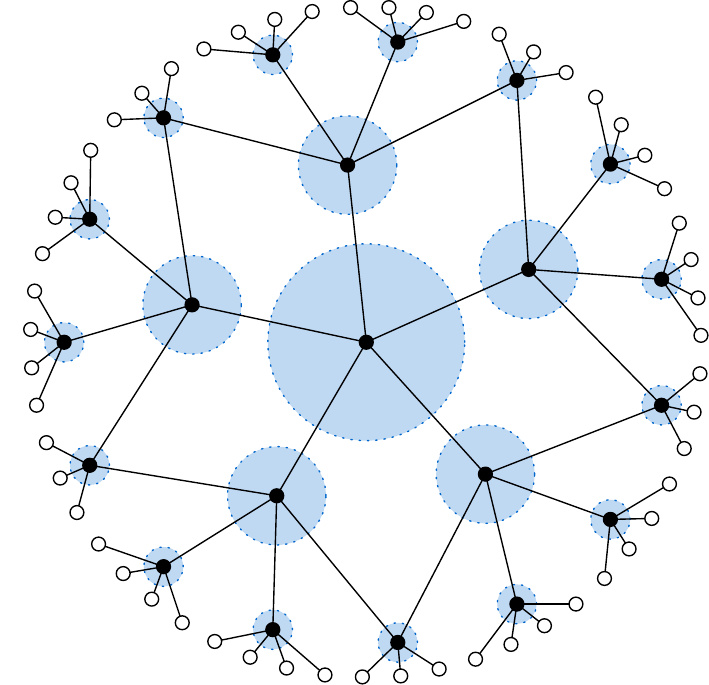}
\caption{Holographic pentagon code. The pentagon tiling of AdS defines a network of negative curvature. Each vertex represents a perfect tensor (indicated by blue disks) in the bulk that takes one qubit as input (represented as black dots). The boundary contains the outputs of the network (represented by white dots). The network of perfect tensors establishes an isometry from the bulk Hilbert space to the boundary Hilbert space and provides a toy model for the AdS/CFT correspondence.}
\label{fig:pentagon_net}
\end{figure}
The ideas outlined in the previous section led to the investigation of extended networks of concatenated quantum error-correcting codes \cite{Pastawski:2015qua, Hayden:2016cfa}. Here, we restrict ourselves to the holographic pentagon code, see figure \ref{fig:pentagon_net}, introduced as a toy model for AdS/CFT in \cite{Pastawski:2015qua} and briefly outline some of the ideas behind its construction.

The basic building block  of the networks of \cite{Pastawski:2015qua} are perfect tensors. These are defined as tensors $T_{a_1,a_2,\dots, a_{2n}}$ with the property that they are proportional to  isometric tensors from $A$ to $A^c$ for all subsets $A$ of the tensor indices with $|A|\leq |A^c|$. In particular, perfect tensors are related to quantum states of $2n$ $v$-dimensional spins as 
\begin{equation}\label{perfectstate}
|\psi\rangle=\sum_{a_1,a_2,\dots, a_{2n}}T_{a_1a_2\dots a_{2n}}|a_1a_2\dots a_{2n}\rangle\,.
\end{equation}
These states $|\psi\rangle$ have the special property that they are maximally entangled along any possible bipartition into sets of $n$ spins and therefore show a very particular entanglement structure.
Specifically, states of this kind are referred to as absolutely maximally entangled states \cite{Helwig:2012nha} and possess interesting properties \cite{Goyeneche:2015fda}.
 Interpreted as a map from one spin to the remaining $2n-1$ spins, a perfect tensor establishes the encoding map of a quantum error-correcting code. It encodes one logical spin into $2n-1$ spins and allows the recovery of the logical one even if up to $n-1$ spins are lost. One explicit example for a perfect quantum error-correcting code that gives rise to a state of the kind described in \eqref{perfectstate} is given by the five qubit code in \cite{Laflamme:1996iw}. The qutrit code described in the previous section provides a further example.

For the construction of the holographic pentagon code, the key idea is to uniformly tile AdS space with pentagons. This tiling defines a network with perfect tensors at each vertex, see figure \ref{fig:pentagon_net}. This tensor network describes an isometric tensor from the bulk (the inputs of the tensor network) to the boundary (its output) and can be seen as a quantum error-correcting code that maps the logical qubits in the bulk to the physical qubits on the boundary. Interestingly, in this network the lattice RT formula holds (see \eqref{latticeRT}). Furthermore, the representation of bulk logical operators on different regions of the boundary is analogous to the reconstruction of bulk operators from CFT operators on the boundary. In consequence, this model captures these important features of the AdS/CFT correspondence.

\section{Classical holographic codes}\label{Classical holographic codes}

In this section, we introduce {\it classical holographic codes}. These are constructed similarly to the holographic quantum error-correcting codes considered in \cite{Pastawski:2015qua}. Spacetime with nonnegative curvature is uniformly tiled. Connecting neighboring tiles we define a network of probabilistic maps. Furthermore, we impose certain constraints on these, as described in section \ref{fullnetwork}. We mainly focus our attention to a network with pentagon symmetry and use bits as bulk and boundary degrees of freedom. However, there are many different constructions possible using, for example, different tilings or trits instead of bits.  It is possible to think about the whole network as a classical error-correcting code. However, we do not refer to our construction as an error-correcting code.\footnote{There are probabilistic codes used for error correction especially in telecommunication. The most prominent examples are low density parity check codes \cite{Gallager} and turbo codes \cite{berrou1995error}. However, there is no straightforward connection between these and the classical holographic codes as we define them here.}  Besides introducing classical holographic codes, we also discuss their features and find some similarities with expectations from AdS/CFT. In particular, we elaborate on close similarities with quantum error-correcting codes that have recently been considered as toy models for AdS/CFT \cite{Almheiri:2014lwa, Pastawski:2015qua, Hayden:2016cfa, Harlow:2016vwg}.

\subsection{Classical trit example}\label{simpleexcl}

To start our discussion on classical holographic codes, we introduce  a 
classical probabilistic code that resembles key features of the quantum case 
discussed in \ref{simpleexqu}. Similar to this case, we consider an encoding of a 
logical trit into three physical ones. Furthermore, we require that the 
information about the logical trit is zero  in each of the individual physical 
trits, while the knowledge of two of the physical trits provides us with full 
knowledge about the logical one. One particular code satisfying these 
constraints is
\begin{equation}\label{tritencoding}
\begin{split}
\tilde{0}\to& \, p(000)=p(111)=p(222)=\frac{1}{3}\,,\\
\tilde{1}\to& \, p(012)=p(120)=p(201)=\frac{1}{3}\,,\\
\tilde{2}\to& \, p(021)=p(102)=p(210)=\frac{1}{3}\,,
\end{split}
\end{equation}
where $p(X_1X_2X_3)$ denotes the probability that the trit string $X_1X_2X_3$ ($X_i\in\{0,1,2\}$) appears. In the encoding (\ref{tritencoding}), each of the strings has the same probability, given by $\frac{1}{3}$.  That is, we encode one logical trit in three physical trits in such a way that the logical one is mapped to three different strings of three trits with equal probability. One can convince oneself that the knowledge of one physical trit does not give any information about the logical one, while by knowing any two physical trits we can obtain the logical one with certainty. Labelling the physical trits by $A$, $B$ and $C$, as above, that implies that the logical trit can be obtained from either $AB$, $AC$ or $BC$, but not from $A$, $B$ or $C$ alone.\footnote{Therefore, codes like the one given by (\ref{tritencoding}) can be used for secret sharing, as we discuss in more detail in section \ref{secret sharing}.} That establishes a subregion duality analogous to the one in the quantum case. 

These properties are also reflected in the Shannon entropy $S_S$. For any of the physical trits $I$, the entropy is given by
\begin{equation}\label{tritentropy}
S_S(I)=-\sum_i p_i \log(p_i)=\log(3)\,,
\end{equation}
where $I=A,B,C$ and the $p_i$ are given by the respective marginal probability distributions. That implies that there is no information about the logical trit in any of the physical ones, as we stated above. Considering any of the sets $AB$, $AC$ or $BC$, we find
\begin{align}
S_S(IJ)=&-\sum_{ij} p_i \tilde{p}_j \log(p_i \tilde{p}_j )\nonumber\\
=&\log(3)+S_S(\tilde{I}) \label{tritentropy2}\,,
\end{align}
where $I,J=A,B,C$, the $p_i$ are the probabilities appearing in (\ref{tritencoding}) and  the $\tilde{p}_j$ give the probabilities for the logical trit $\tilde{I}$ to be $\tilde{X}$ ($\tilde{X}\in \{\tilde{0},\tilde{1},\tilde{2}\}$) and we used $\sum_i p_i=1=\sum_j \tilde{p}_j$. First, we notice that these results are formally the same as in the quantum case discussed in \ref{simpleexqu}. That is, a RT formula -- at least formally -- holds. However, the RT formula is concerned with entanglement entropy, while here we considered the Shannon entropy. To connect both, we move to the mutual information that, for pure states, is equal to two times the entanglement entropy. We find that the mutual information $I_{cl}$ between one physical trit $A$ and the remaining two is given by
\begin{equation}
I_{cl}(A,BC)= S_S(A)+S_S(BC)-S_S(ABC)=\log(3)\,,
\end{equation}
where we used $S_S(BC)=S_S(ABC)=\log(3)+S_S(\tilde{I})$. Because of the symmetry of the encoding the same statement also holds for the other two trits $B$ and $C$. That is, the classical mutual information is smaller than the one in the quantum case, (\ref{quantummutinfo}), by a factor of $\frac{1}{2}$. However, it also is proportional to the ``area'' of the minimal cut.

Let us next investigate whether we can implement logical operations in the bulk ({\it i.e.}, on the logical trit) by acting on a subset of the boundary degrees of freedom (the physical trits), see figure \ref{qutrit}. First, let us implement an operation that implements addition by $\oplus 1$   by solely acting on the physical trits $B$ and $C$.\footnote{Here and in the remainder of this section, $\oplus n$ for some integer $n$ denotes the addition by $n$ mod$3$.} The operation that succeeds in this task is to apply $\oplus 1$  to $B$ and  $\oplus 2$  to $C$. The same operation can be implemented on $A$ and $B$ by applying $\oplus 1$  to $A$ and  $\oplus 2$  to $B$. Finally, to implement it on $A$ and $C$, one has to apply $\oplus 2$ to $A$ and  $\oplus 1$  to $C$. To perform the logical operation $\oplus 2$ by acting on two of the physical trits, one has to either act with $\oplus 2$ on $B$ and $\oplus 1$ on $C$, with $\oplus 2$ on $A$ and $\oplus 1$ on $B$ or with $\oplus 1$ on $A$ and $\oplus 2$ on $C$. Therefore, operators acting on the logical trit can be reconstructed on either $AB$, $AC$ or $BC$, but not on $A$, $B$ or $C$ alone.

In summary, the classical code we considered shares essential features with the quantum code that we reviewed in section \ref{simpleexqu}.

Furthermore, it is interesting to note that the encoding (\ref{tritencoding}) can be obtained from (\ref{qutritencoding}) by imposing complete decoherence.\footnote{Note that the classical encoding (\ref{tritencoding}) does not have to be obtained in this way nor does it have to be interpreted in this way. Also, already at this point, we want to mention that the classical codes on extended networks, which we introduce in the next section, cannot be obtained by decoherence of the boundary state of \textit{e.g.} the holographic pentagon code.} Mapping the classical logical trit given by $\tilde{i}$ ($\tilde{i}\in\{\tilde{0},\tilde{1},\tilde{2}\}$) to the logical qutrit state $|\tilde{i}\rangle$ and subsequent encoding according to (\ref{qutritencoding}), we obtain 
\begin{equation}
\rho_{\tilde{i}}=|\tilde{i}\rangle\langle  \tilde{i}|=\frac{1}{3}\left(\begin{array}{cccc}
1&1&1&0_{1\times 6}\\
1&1&1&0_{1\times 6}\\
1&1&1&0_{1\times 6}\\
0_{6\times 1}&0_{6\times 1}&0_{6\times 1}&0_{6\times 6}
\end{array}\right)
\end{equation}
in a basis containing the qutrit states appearing in \eqref{qutritencoding}, where we denote the basis by $\{|v_{j}\rangle\}_{j=1,\dots,9}$, and the ordering depends on $\tilde{i}$. Removing the coherences in $\rho_{\tilde{i}}$, for example, by a randomly selected projective measurement with projectors $P_j=|v_j\rangle\langle v_j|$, we arrive at a mixed state $\rho_{\tilde{i}}^{(dec)}=\frac{1}{3}\sum_{j=1}^3 |v_j\rangle\langle v_j|$. This is a statistical mixture of pure states $|v_j\rangle\langle v_j|$ that appear with probability $p(v_j)=\frac{1}{3}$. Therefore, by reinterpreting the qutrits as classical trits, we obtain the encoding (\ref{tritencoding}). 

At this point, we would like to insert another brief comment. There is the question how the randomness in the description of the system can be justified physically. In our opinion, there are (at least) three possible ways. One is that there is a lack of knowledge about the details of the system that forces a probabilistic description, as in thermodynamics (cf.\ section \ref{Interpretation}). Another way to justify the randomness in the  code is to imagine an agent at each vertex that generates the randomness that is necessary for the functioning of the code, for example, by sending individual photons to a beam splitter and subsequently collapsing the quantum superposition of the photons. In this way the agent can create the required random numbers. Similarly, one could think of strong local decoherence at each of the vertices that kills the coherences and leaves us with a probabilistic mixture, as described above. However, in our opinion, it also is enough to just state that the codes we consider are intrinsically random.

\subsection{Classical codes on hyperbolic space}\label{fullnetwork}

\begin{figure}[t]
 \centering
 \includegraphics[width=8cm]{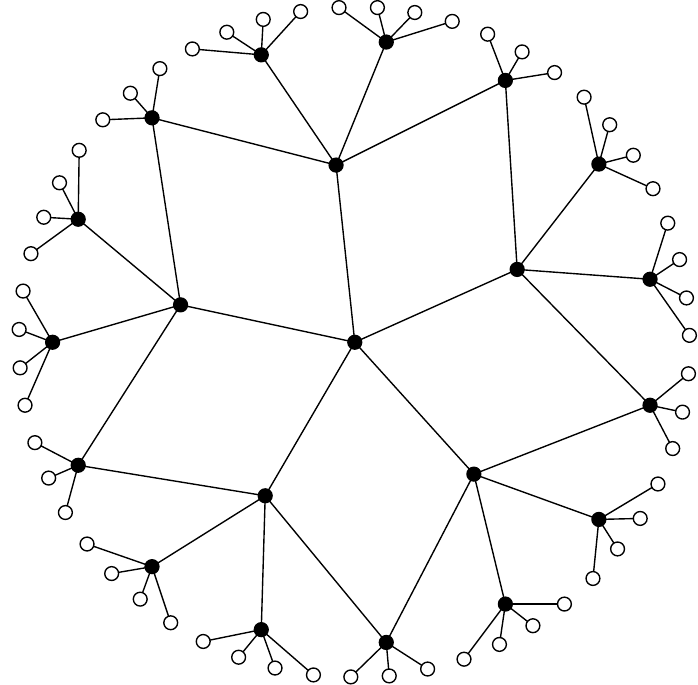}
 \caption{Network to realize a classical holographic code. Each vertex in the interior of the graph represents a tile with a specific fixed volume in AdS space. Furthermore, each of these vertices takes one bit as input (the input bits are then interpreted as bulk degrees of freedom) and (probabilistically) maps the input together with the input from the ingoing edges  to the outgoing edges. The final output of the code is then given by the bits sitting at the boundary of the network. These are interpreted as boundary degrees of freedom. In this way a map from bulk degrees of freedom to boundary degrees of freedom is established that gives rise to a duality between bulk and boundary.}
 \label{fig:network}
\end{figure}

We study classical probabilistic codes on a uniform pentagon tiling of AdS space that feature some of the key properties of tensor-network-based quantum codes \cite{Pastawski:2015qua, Harlow:2016vwg, Hayden:2016cfa, Almheiri:2014lwa} under which there are the Ryu-Takayanagi formula and important bulk reconstruction properties. The tiling gives rise to a network, that we also refer to as graph, as \textit{e.g.} visible in figure \ref{fig:network}. Via the network, we define a (probabilistic) mapping from the bits sitting on the vertices in the interior to those on the open edges at the boundary. The mapping is defined as follows: We order the network into layers of vertices defined by the graph distance from the center. From the negative curvature of the graph it follows that each vertex shares at most two edges with vertices of the previous layer. We now declare each node to a map $n \rightarrow m$, where $n$ is the number of inputs given by the bit at the vertex and edges from the previous layer, and $m$ is the number of output bits. There are three possible mappings appearing in this pentagon tiling, shown in Figure \ref{fig:network}, are  $3\rightarrow3$, $2\rightarrow4$, and, in the center, $1\rightarrow5$.

\begin{figure}\centering
\includegraphics{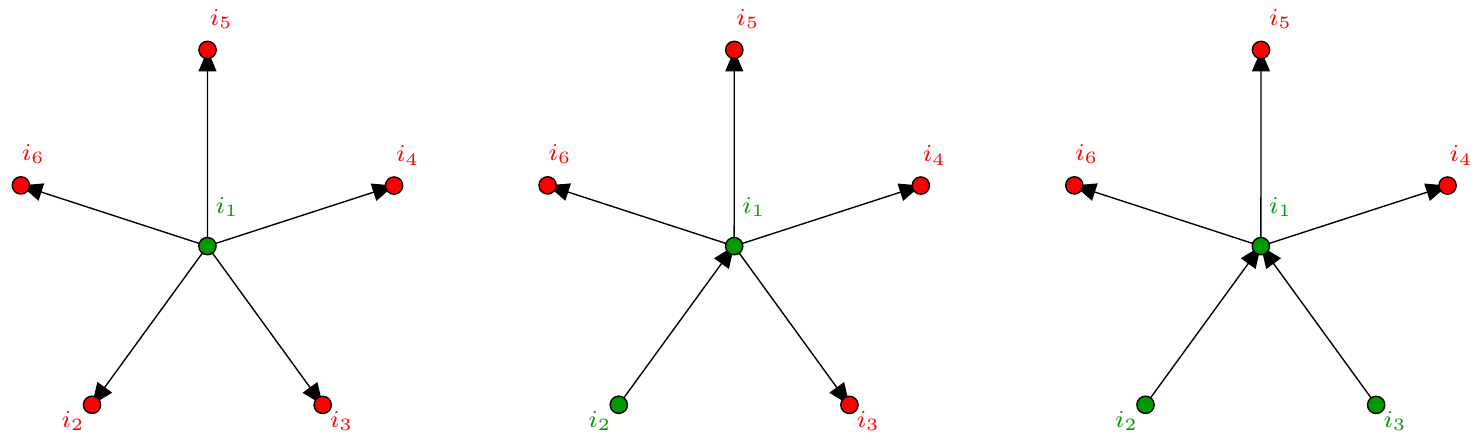}
 \caption{The set $\mathfrak{S}= \{s_i\}$ of strings $s_i$ of bits $i_n$ generates the (probabilistic) mappings. Here, we display from the left to the right the mappings: $1\rightarrow5$, $2\rightarrow4$, and $3\rightarrow3$. Each dot represents one bit, where the one in the center, $i_1$, is the bulk input and the remaining ones are edge input (green) and outputs (red). The probability distribution of the outputs is obtained via the conditional probabilities $ p_\text{out}(s_\text{out}\,|\, s_\text{in})$, where $s_\text{in}$ denotes the string of inputs and $s_\text{out}$ is the string of outputs.}\label{fig:mappings}
\end{figure}

Inspired by the quantum codes of \cite{Pastawski:2015qua}, where the mapping from the bulk to the boundary is due to the insertion of one and the same perfect six-tensor at each vertex, we demand that each mapping originates from a single set of strings of six bits, $\mathfrak{S} = \{s_i\,|\, i=1,..,N\}$, with the number of strings, $N$, not yet fixed. We define the mapping as follows: The first bits in the strings define the input, where we always take the very first bit in the strings as the bulk input (see fig. \ref{fig:mappings}). We now assume a discrete uniform probability distribution on the set $\mathfrak{S}$, \textit{i.e.}, all probabilities $p(s_i) = 1/N$ are equal. The probability density of the outcome of the mappings for a given input string $s_\text{in}$ is then defined by the conditional probabilities

\begin{equation}
 p_\text{out}(s_\text{out}\,|\, s_\text{in})\,, 
\end{equation}
where $s_\text{in}\cup s_\text{out}\in \mathfrak{S}$. The domain of the $3\rightarrow3$ mapping should contain all possible strings of three bits. This gives a first condition on $\mathfrak{S}$ and tells us that $|\mathfrak{S}| = N \ge 8$.

As we discussed in section \ref{Pentagon_code}, a perfect tensor gives rise to an absolutely maximally entangled state. In \cite{Pastawski:2015qua}, this particular entanglement structure was used to show the desired features.  Here, we demand rather similar conditions for  $\mathfrak{S}$, where we  use the mutual information as a measure of correlations. As it turns out, it is not possible to find a set $\mathfrak{S}$ of bit strings of length six, where any bipartition of the strings has maximal mutual information, which in a sense would be the classical analogue to perfect tensors.  These analogues exist only for some special combinations of the length of the strings and the ``dimension" $d$ of the $d$its involved. It is not clear whether such a set of strings of length six exists. Therefore, in the following, we prefer to use milder conditions on the set $\mathfrak{S}$ that still will be sufficient to obtain the results of section \ref{features}. The same properties then follow automatically for sets of maximally correlated strings. We demand that any bipartition into substrings of non equal size is maximally correlated, and that any bipartition into strings of length three is maximally correlated if one of the two substrings contains only neighboring bits. 
Here, the term neighboring bits refers either to bits that are next to each other in the full (cyclic) six-string or to bits where the edges, that  are allocated to these, are next to each other (see fig. \ref{fig:mappings} for the allocation). Therefore, the order in which the bits appear in the string and whether a particular bit acts as edge in- or outputs matter. As illustrated in figure \ref{fig:mappings}, we choose the bits to be arranged counterclockwise.

As we show in appendix \ref{properties}, from the above properties, it follows that $|\mathfrak{S}| =  8$ and that
\begin{itemize}
\item[(I)] the knowledge of three neighboring edge bits gives full information about the three complementary bits;
\item[(II)] no information about any other single bit can be obtained by the knowledge of one particular bit. 
\end{itemize}
Furthermore, two neighboring edge bits never reveal information about bits next to them and in general two bits can at most give one other bit with certainty.

\definecolor{ffqqqq}{rgb}{1.,0.,0.}
\definecolor{qqzzqq}{rgb}{0.,0.6,0.}
After this more general discussion, we give an explicit example of a set of strings that fulfills the above properties. It is given by
\begin{equation}\label{examplesetofstrings}
 \mathfrak{S} = \{000000, 001111, 010110, 011001, 100101, 101010, 110011, 111100\}\,,
\end{equation}
with the probability distribution $p(s_i\in \mathfrak{S}) = \frac{1}{8}$. In consequence, the  $1\rightarrow5$ mapping is given by
\begin{equation}
\begin{split}
\tilde{0}\to&~ p(00000)=p(01111)=p(10110)=p(11001)=\frac{1}{4}\,,\\
\tilde{1}\to&~ p(11100)=p(10011)=p(00101)=p(01010)=\frac{1}{4}\,, 
\end{split}\label{1to5encoding}
\end{equation}
where here and in the following the tilde indicates the bulk input and $p(X_1X_2X_3X_4X_5)$ denotes the probability of the output $X_1X_2X_3X_4X_5$ ($X_i\in\{0,1\}$). Unfortunately, the mapping breaks the pentagon symmetry of the network. This is because  the central bulk bit can be  reconstructed with the knowledge of the second and fifth or the third and fourth output bits but not with the knowledge of any other two bits. Therefore these bits are distinguished. All sets $\mathfrak{S}$ give rise to $1\rightarrow5$ mappings that break the symmetry in a similar way. However, this does not spoil the desired property for the full network. 

For  $2\rightarrow4$ we obtain
\begin{equation}
\begin{split}
\tilde{0}0_{e}\to&~ p(0000)=p(1111)=\frac{1}{2}\,,\quad
\tilde{1}0_{e}\to p(0101)=p(1010)=\frac{1}{2}\,,\\
\tilde{0}1_{e}\to&~ p(0110)=p(1001)=\frac{1}{2}\,,\quad
\tilde{1}1_{e}\to p(1100)=p(0011)=\frac{1}{2}\,,
\end{split}\label{2to4encoding}
\end{equation}
where the subscript $e$ indicates the edge input from the previous layer. Finally, the $3\rightarrow3$ map deduced from the set (\ref{examplesetofstrings}) is given by 
\begin{equation}
\begin{split}
\tilde{0}0_{e_1}0_{e_2}\to&~   000 \,, \quad \tilde{0}0_{e_1}1_{e_2}\to   111 \,, \quad \tilde{1}0_{e_1}0_{e_2}\to   101 \,, \quad \tilde{1}0_{e_1}1_{e_2}\to   010 \,, \\
\tilde{0}1_{e_1}0_{e_2}\to&~   110 \,, \quad \tilde{0}1_{e_1}1_{e_2}\to   001 \,, \quad \tilde{1}1_{e_1}0_{e_2}\to   011 \,, \quad \tilde{1}1_{e_1}1_{e_2}\to   100 \,, 
\end{split}\label{eq:3to3}
\end{equation}
where $e_1$ and $e_2$ denote the bits of the incoming edges. 

In the following, we show that a map from the bulk to the boundary induced by  a set with the outlined properties -- and in particular the specific example (\ref{examplesetofstrings})  -- together with the geometric structure of the network inherit the above mentioned features. For that reason, we call them \textit{classical holographic codes}. In particular, the properties we demand on $\mathfrak{S}$ are sufficient to obtain the results of the next section.

The above approach is a generic way to construct codes on a hyperbolic space that give rise to the features we show in the following section. However, there are many more possible probabilistic codes that work, too. One can \textit{e.g.} define each individual map by a different set that fulfills the above property of maximal mutual information. This does not alter any property we study in section \ref{features}. One can also consider the situation where the maximally correlated set that defines the mapping is random at each vertex. In this case the classical version of the RT formula still holds.\footnote{A reconstruction of the bulk degrees of freedom is no longer possible, as for this task the knowledge of the mapping at each vertex is required. For fixed (and therefore known) mappings at each vertex that are obtained by sampling from some probability distribution, however, all properties we obtain in  \ref{features} still hold.} This establishes a similarity with the random tensor networks considered in \cite{Hayden:2016cfa}.

In particular, we emphasize that, in contrast with the simple example of section \ref{simpleexcl}, the classical codes introduced in this section cannot be obtained by simple decoherence of the boundary. That is, decoherence of the output of a quantum code, as \textit{e.g.} the one considered in section \ref{Pentagon_code}, does lead to a different probability distribution. In particular, it is not at all clear why the resulting system should possess any special properties. In general, that is surely not the case.

\subsection{Features of classical holographic codes}\label{features}

In this section, we investigate to what extent classical probabilistic codes defined by a network on AdS produce properties similar to those of quantum error-correcting codes. As we find, classical holographic codes possess several interesting properties that are analogous to properties of QECC and, in particular, AdS/CFT.

\subsubsection{Ryu-Takayanagi formula}

Consider a CFT with a gravitational dual, where at least for every static state at low energies there exists a geometric bulk description. In these states, the Ryu-Takayanagi (RT) formula\footnote{Here and in the following, we do not consider  contributions from bulk entropy.} relates the entanglement entropy $S_A$ of a boundary region $A$ at fixed time to the area of the minimal surface $\gamma_A$ in the bulk, whose boundary coincides with the boundary of $A$
\begin{equation}
 S_A = \frac{\text{Area}(\gamma_A)}{4 G}\,,
\end{equation}
where $G$ is Newton's constant \cite{Ryu:2006bv, Ryu:2006ef}. 

An analogous relation holds for the quantum error-correcting codes considered in \cite{Pastawski:2015qua, Harlow:2016vwg}. Considering a so-called holographic state -- that is a boundary state of a tensor network of perfect tensors with a graph of nonpositive curvature -- then measured in units of $\log(2)$ the entanglement entropy of any connected region $A$ on the boundary equals the length of the shortest cut\footnote{A cut is a path through the network that separates it into two disjoint sets of vertices and the length of the cut is given by the number of edges it crosses.} $\gamma_A$ through the network whose boundary matches that of $A$
\begin{equation}\label{latticeRT}
 S_A = |\gamma_A|\,.
\end{equation}
That is, for these tensor networks, the lattice RT formula holds. 

Interestingly, in the case of a classical holographic code a very similar statement is true. Of course, the concept of entanglement entropy does not exist in classical systems. In particular, there is no quantum entanglement. However, if we interpret this quantity not only as a measure of quantum entanglement but of correlations, or even more abstract as a measure of joint information between two subsystems, then there is a classical analogue namely the mutual information $I_{cl}$. It can formally  be defined in the same way for both classical and quantum theories 
\begin{equation}
 I_{qu/cl}(A,B) = S(A) + S(B) - S(A,B) \,,
\end{equation}
where $A$ and $B$ denote two subsystems and the subscripts $qu$ and $cl$ specify the quantum mutual information $I_{qu}$, defined in terms of von Neumann entropies, and the classical mutual information $I_{cl}$, defined in terms of Shannon entropies. In a quantum theory, $S(A)$ and $S(B)$ are the von Neumann entropies of the respective reduced density matrices of subsystems $A$ and $B$. $S(A,B)$ denotes in this case the von Neumann entropy of the union of $A$ and $B$. For a bipartition of a system in a pure state into $A$ and $B = A^c$, the total entanglement entropy vanishes and the two partitions show equal entropy, $S(A) = S(B)\equiv S_A$, such that 
\begin{equation}\label{mutinfopurestate}
 I_{qu}(A,A^c) = 2 S_A\,.
\end{equation}
In a classical system $S(\cdot) \equiv S_S(\cdot)$ denotes the Shannon (or marginal) entropy of the system inside the bracket. As in the quantum case, the mutual information measures the joint information of the two subsystems $A$ and $B$. However, for classical systems, the mutual information is solely due to classical correlations between subsystems. 

The considerations above lead us to the conclusion that the mutual information is the natural candidate to quantify classical correlations between distinct parts of the classical system of interest. Further motivation to single it out as the measure of correlations in the present work is provided by its close relation to the entanglement entropy for pure states given in (\ref{mutinfopurestate}). Therefore, in what follows we formulate and proof a formula in terms of the mutual information that shows the same behavior as the lattice version of the RT formula for QECCs. The  intuition behind this formula is that it establishes a duality between a geometric quantity in the bulk -- namely the length of the minimal cut -- and classical correlations, as measured by the mutual information, on the boundary. This is closely analogous to the RT formula in AdS/CFT, where, for pure boundary states, the statement is that the entanglement entropy given by half of the mutual information is proportional to the area of a minimal surface in the bulk.

{\bf A version of the RT formula for classical holographic codes}\hspace{3mm}  For an arbitrary but fixed bulk input, the classical mutual information between a (connected) subregion $A$ on the boundary and its complement $A^c$ is given by the length of the minimal cut $\gamma_A$ through the network, whose boundary matches that of $A$, 
\begin{equation}\label{classicalRTformula}
 I_{cl}(A,A^c) = |\gamma_A|\,.
\end{equation}
Therefore, a version of the RT formula holds for these classical systems. The length of the minimal cut equals classical correlations on the boundary.\footnote{Note that the lattice RT formula (\ref{latticeRT}), that was proven for holographic quantum error-correcting codes, can, for pure boundary states, be written in terms of the mutual information, as $ I_{qu}(A,A^c) = 2 |\gamma_A|$. Thus, it is evident that for quantum codes the mutual information is twice the classical one.}

The proof of (\ref{classicalRTformula}) that we give in appendix \ref{proof} proceeds along the following steps. First, we argue that the length of any cut dividing the network into two parts provides an upper bound for the mutual information. Therefore, it is clear that the minimal cut $\gamma_A$ gives the smallest upper bound. Subsequently, we complete the proof by showing that the edges that are crossed by the cut are uncorrelated. From that, it follows that the bound is saturated and, thus, (\ref{classicalRTformula}) holds.

\subsubsection{Bulk and operator reconstruction}

\begin{figure}[t]
 \centering
\includegraphics{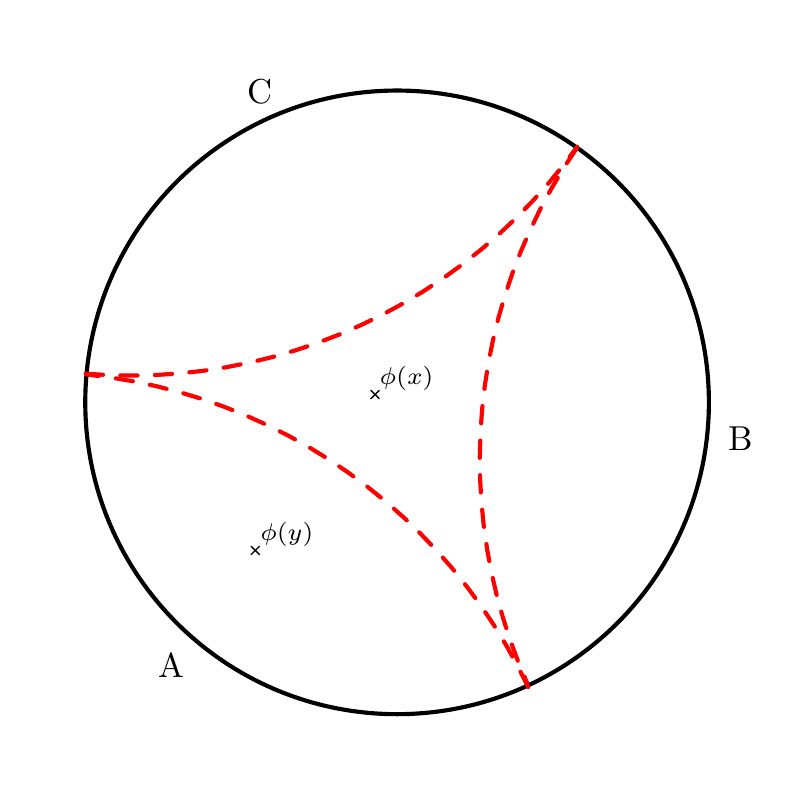}
\vspace{-.8cm}
 \caption{Subregion duality. The operator $\phi(y)$ can be represented on the boundary region $A$, but not on the complementary region $A^c=BC$. The operator $\phi(x)$, however, cannot be represented on any region $A$, $B$ or $C$. But, still, it can be represented on the union of any two of these regions, {\it  i.e.}, on $AB$, $AC$ and $BC$. That is referred to as subregion duality.}
 \label{subregionduality}
\end{figure}

In AdS/CFT, a gravitational theory on $(d+1)$-dimensional asymptotically AdS space (bulk) is related to a $d$-dimensional conformal field theory on the boundary. That immediately raises the question how, given some configuration of the boundary, the bulk can be reconstructed. This is, in particular, complicated by the emergent spatial dimension. The information required to reconstruct some region of the bulk is contained in a boundary region if its entanglement wedge contains this region of the bulk \cite{Dong:2016eik}, see figure \ref{subregionduality}.  Here, in the classical case, we argue that the relevant wedge is the {\it correlation wedge} $C(A)$ that is defined as the region bounded by the minimal cut. It is therefore very similar to entanglement wedge reconstruction. In the following, we demonstrate the possibility of bulk reconstruction in the {\it correlation wedge} of a region of the boundary. Furthermore, we address the issue of operator reconstruction  and show that -- in our case -- classical operations, like bit flips, on bulk degrees of freedom contained in the correlation wedge of some boundary region $A$ can be performed by acting (nonlocally) on the boundary degrees of freedom in $A$.

Let us assume $A$ is connected and the minimal cut $\gamma_A$ is unique, and then we can reconstruct every bulk input bit in $C(A)$. This is evident by considering the algorithm for constructing the minimal cut as described in appendix \ref{proof}. In every step, it crosses three neighboring edges that allow one to reconstruct all the other bits, including the bulk input of the vertex it jumps over -- due to property (I). 

Most bulk inputs in the complement of $C(A)$ cannot be reconstructed with some exceptions. These occur when the minimal cut crosses two neighboring bits from a vertex outside of the correlation wedge. Then the conditions we demand for the code  allow that, for example, these two edge bits are maximally correlated  with the respective bulk input and, hence, it can be reconstructed. This is visible in our explicit example and is most evident if we consider the $2\to4$ mapping given in \eqref{2to4encoding}. If the minimal cut crosses the second and third outputs, their knowledge immediately allows one to reconstruct the bulk input. Besides these cases, that only allow one to reconstruct inputs directly behind the minimal cut, no other bulk inputs in the complement of $C(A)$ can be reconstructed. We do not consider the exceptions as a crucial problem, as in the limit of large networks, \textit{i.e.}, where the number of bulk inputs goes to infinity, this effect is negligible.

\begin{figure}[ht]
 \centering
 \includegraphics[width=12cm]{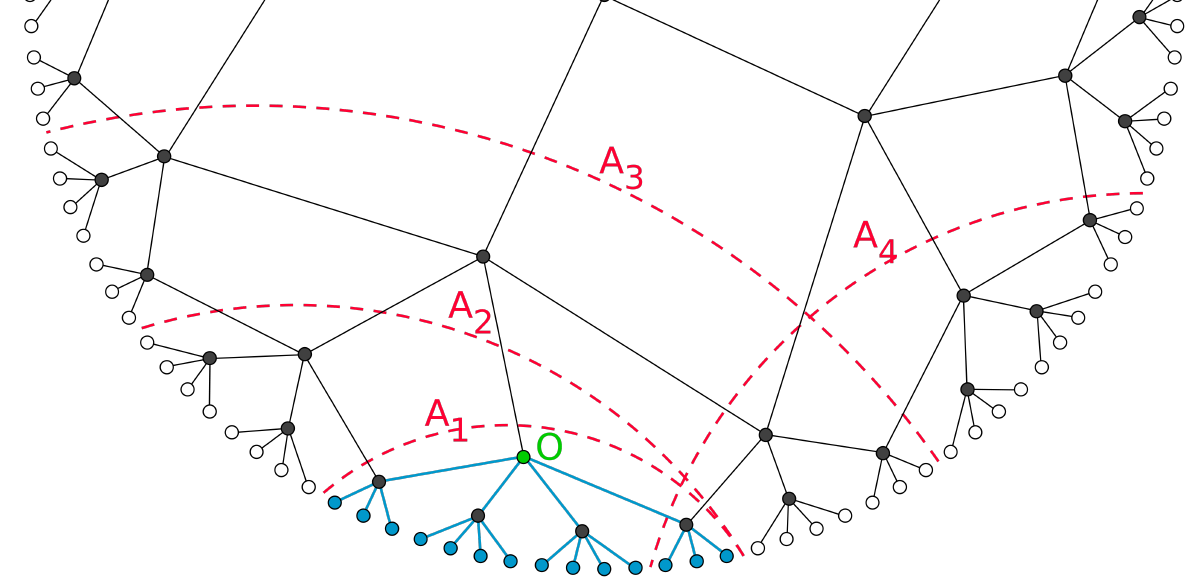}
 \caption{Representing bulk operations. The action of an operation $O$ on one of the bulk bits, bit $I$, can definitely be represented on a boundary region $A_i$ if $I$ is contained in $C(A)$. Here, we show one particular example and marked the edges and vertices blue that can be affected by the operation on the green vertex. Here, $C(A_{1,2,3})$ contain the bit $I$ and hence the bit flip $O$ can be realized on these boundary regions. $A_4$ is an example that does not allow to reconstruct $O$.}
 \label{fig:oprecon}
\end{figure}

Next, we consider the reconstruction of bulk operations.\footnote{Note that in a classical code the ``bit flit operator'' $O$ is the only nontrivial operation.} Assuming a connected boundary region $A$, all bit flip operations $O$ on vertices in the bulk region $C(A)$ can be represented as multiple bit flips in $A$. The reason for this is the following. From the algorithm to construct the minimal cut as given in appendix \ref{proof} it follows that any vertex in region $C(A)$ has at least three neighboring edges that are contained in $C(A)$ and go in the direction of $A$; see figure \ref{fig:oprecon}. Solely flipping some of these bits cannot affect bits in the complement of $A$. Therefore, degrees of freedom in $A$ are sufficient to reconstruct operations in $C(A)$. Consider now the action of an operation $O$ on a vertex in $C(A)$. Then it is possible to successively modify the edge bits in $C(A)$ until we reach the boundary region $A$. Obviously, no edge bit leaving $C(A)$ is touched by this procedure. Therefore the operation $O$ on any bulk bit in $C(A)$ can be reconstructed by flipping the respective subset in $A$ that was flipped by the above procedure. This is, in general, not possible for bit flips on vertices in the complement of $C(A)$.

Another question arising is whether the operations on the boundary region that realize a specific bit flip in the bulk depend on the configuration of the boundary bits. For the example given in \eqref{1to5encoding}, \eqref{2to4encoding}, and \eqref{eq:3to3} this is not the case. This becomes evident if we look at the individual mappings. Flipping some inputs in a specific way always leads to the same possible flips in the output, independent of the actual values of the bits. For example, flipping the bulk input in the $3\to3$ map always flips the first and third output bits, or solely changing the edge input in the $2\to4$ mapping can always be realized by changing the second and third outputs. It never depends on the actual value of the bits. This holds for any mapping in the network, so in total it holds for the entire network. Therefore, the boundary realization of a bit flip operation on some bulk bit does not depend on the boundary configuration.\footnote{At this point, it might be worth emphasizing a difference of our code and tensor network models. In tensor network models, in general, the boundary representation of an operator is a complicated nonlocal object, while, here, the boundary operation is a ``tensor product'' bit flip operation.} However, it is in general not unique. For a flip operation on one of the inputs, the $1\to5$ and the $2\to4$ mappings allow different realizations on the output. In our example in \eqref{2to4encoding}, a bulk input flip can be realized by flipping the first and third outputs or by flipping the second and fourth outputs. In general, a flip in the bulk has more representations on the boundary the deeper in the bulk it is located.

{\bf Subregion duality}\hspace{3mm} The so-called subregion duality in AdS/CFT states that operators in the bulk can, in general, be represented on different subregions of the boundary, see figure \ref{subregionduality}. In \cite{Almheiri:2014lwa}, the toy model we reviewed in section \ref{simpleexqu} was suggested to capture essential features of this duality. Also in more elaborate tensor network models based on quantum error-correcting codes, it was shown to hold \cite{Pastawski:2015qua}. Here, we show that also in the classical network, we introduced, there is a notion of subregion duality. Indeed, it immediately follows from the fact that an operation $O$ on any bulk input $I$ can be represented on a boundary region $A_i$ if $I\in C(A_i)$, as we have shown above; also see figure \ref{fig:oprecon}. Therefore, all representations of $O$ on each of the $A_i$'s are dual to each other. This establishes a notion of subregion duality for classical holographic codes.

{\bf Black holes}\hspace{3mm} A naive picture of asymptotically AdS spacetimes containing black holes is to describe these configurations by ``cutting out" some region of the network \cite{Pastawski:2015qua}. The microstates of black holes are then described by the edge bits crossed by the horizon that function as inputs for the remaining network. In consequence, the black hole has a nonvanishing entropy that scales as the number of edges crossed by the horizon, {\it i.e.}, it scales as the area of the black hole. Interestingly, this behavior is only expected in the semiclassical approach \cite{Hawking:1974sw,Bekenstein:1973ur} and should not appear at the classical level. However, we emphasize that this picture of black holes is very naive.

\subsubsection{Secret sharing}\label{secret sharing}

Finally, we insert a brief discussion of the secret sharing property of classical holographic codes. The fact that these codes possess this property provides further motivation for their construction beyond the holographic interpretation. Secret sharing codes are characterized by the fact that there is a secret information (some string of bits) -- or secret for short -- that is distributed amongst several parties such that each party individually has no access to the secret. If, however, a sufficient number of parties collaborate, they can gain access to the secret \cite{shamir1979share, blakley1979safeguarding}.

Let us start the discussion by showing that the simple trit example that we introduced in section \ref{simpleexcl}  falls into the class of secret sharing codes. In this example, we view the input trit as the secret. The probabilistic code (\ref{tritencoding}) distributes the secret amongst three parties such that each party gets exactly one trit. Since the Shannon entropy of each of the three trits is maximal, $S_S=\log(3)$, an individual party has no information about the secret. However, as soon as two arbitrary parties collaborate and share their trits, they obtain full information about the secret. Thus, the code (\ref{tritencoding}) is a $(n=3, t=2)$-threshold scheme, where $n$ denotes the number of parties and $t$ denotes the threshold of parties that is necessary to obtain the secret.

As we argue next, classical holographic codes also belong to the class of secret sharing codes. To see this, we interpret the bulk inputs as the secret to be shared. Imagine now that each party is in possession of one of the boundary bits.\footnote{Of course, it does not have to be exactly one of the boundary bits per party, but also larger fractions of the boundary bits can be in possession of each party. However, for the sake of clarity and simplicity, let us assume that situation.} Then, individually, each party has no chance to learn about any of the bulk inputs. However, by collaborating, i.e., by sharing their knowledge of their respective boundary bits, a team (a set of parties) can learn (part) of the bulk inputs (part of the secret). An illustrative example is the setting in which all bulk inputs are publicly known, except for the one in the center. We refer to the center bit as the secret. In this case, once a sufficient number of parties\footnote{Here, a team of roughly more than 50\% of the parties is sufficient.} team up they can reveal the secret, while the remaining ones obtain no knowledge at all about the secret. Therefore, classical holographic codes are secret sharing codes.

\section{On a possible physical interpretation}\label{Interpretation}

While so far we have discussed classical holographic codes and their properties in a rather abstract way, in this section, we give a possible physical interpretation of these. In particular, we focus on the radial spacelike direction and connect it to coarse graining in phase space, where the main idea is to interpret the additional bulk direction as parameter for an effective description of the boundary. This is similar to the interpretation of the radial direction in AdS as geometrizing the renormalization group flow of the dual CFT (see \textit{e.g.} \cite{CasalderreySolana:2011us,Susskind:1998dq}).

In our case, we interpret the boundary degrees of freedom/code subspace as the microstates of a  classical statistical system characterized by a probability distribution in a discretized phase space. To simplify the following considerations, but without loss of generality, we assume the probability distribution to be uniform within its support in phase space. Then the discretization is such that the region of phase space that supports the probability distribution is tiled with tiles of equal volume. It is the bulk inputs in the layer next to the boundary that dictate the support of the distribution, {\it i.e.}, each bulk input corresponds to the location of one of the tiles in phase space.

\begin{figure}[ht]
 \centering
 \includegraphics[width=10cm]{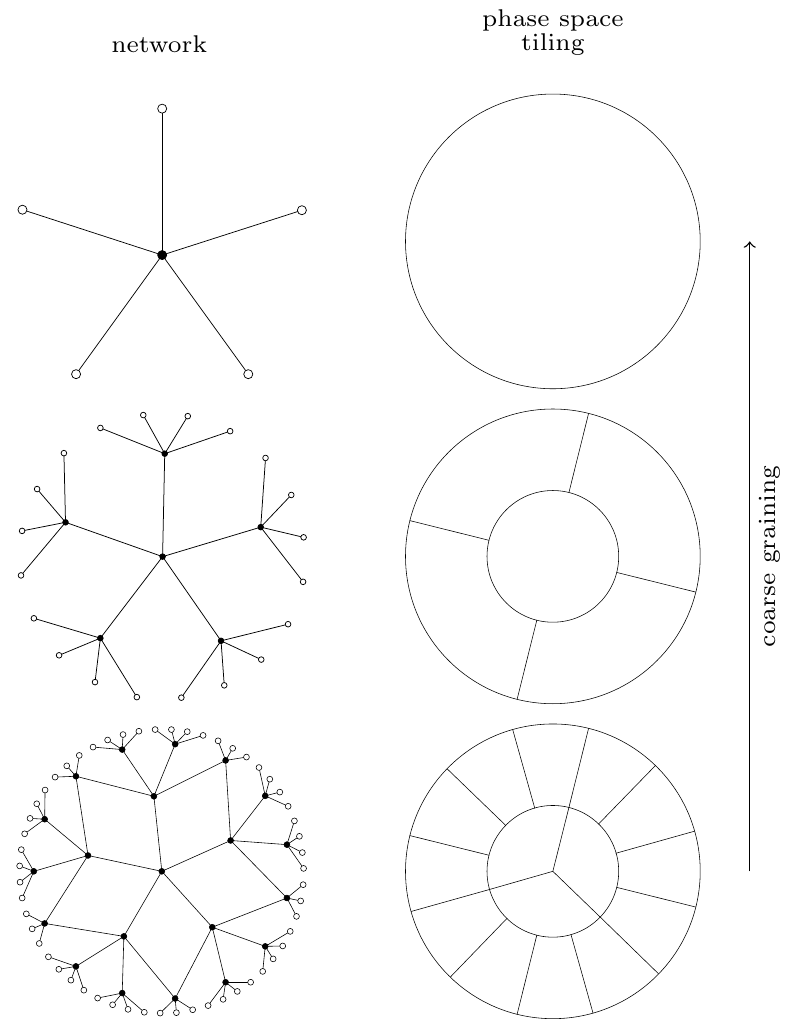}
 \caption{The bulk direction is interpreted as a coarse graining parameter for an effective description of the boundary. It interpolates between the microscopic description at the boundary of AdS and the macroscopic description in the center of AdS, while both are connected by coarse graining phase space.}
 \label{fig:interpretation}
\end{figure}

Then each step in the radial direction, {\it i.e.}, considering the network with one reduced layer, corresponds to joining\footnote{In general, the coarse graining does not necessarily require to join tiles pairwise. In principle, any constant number $k$ of tiles can be joined in each step. $k$ depends on the structure of the underlying graph defined by the classical holographic code.} neighboring tiles and, therefore, by going deeper into the bulk, a more and more coarse grained description of the system is obtained. In terms of bulk inputs, moving inward for one layer of the graph means that the number of bulk inputs in this layer is strictly smaller than the one in the previous layer. The same is true for the number of boundary degrees of freedom. This number also decreases with each step. Therefore, coarse graining naturally emerges, see figure  \ref{fig:interpretation}.

Thus, we interpret the bulk direction as a coarse graining parameter for an effective description of the boundary. It interpolates between the microscopic description at the boundary of AdS and the macroscopic, \textit{i.e.} thermodynamic, description in the center of AdS, while both are connected by coarse graining phase space. From these considerations, the analogy to the renormalization group flow on the CFT side of AdS/CFT becomes apparent. In AdS/CFT the radial direction can be thought of as a geometric manifestation of the renormalization group flow from the UV to the IR fix point.

To illustrate the idea, consider the microcanonical description of a free gas. The probability density $\rho(x,p)$, where $x$ and $p$ denote the position and momentum of the particles, has support only in the close vicinity to the sphere characterized by $E=\sum_i\frac{m_i}{2}|p_i|^2$ in phase space, where $E$ is the total energy and $m_i$ is the mass of particle $i$. We denote the sphere by $S_E$. Then, macroscopically, the system is completely characterized by one (macroscopic) variable, the total energy $E$. In phase space, this can be viewed as maximally ignorant description (in our language, a completely course grained description), where one only cares about the fact that the underlying microscopic state of the system actually is described by an arbitrary point in $S_E$. In case of classical networks, this is the description in the center of the bulk. Let us now consider a more fine grained description, for example, by dividing  $S_E$ in $k$ ($k\in \mathbb{N},k> 1$) patches of equal volume. Physically, the more fine grained description is due to some additional knowledge. For example, one might for some reason be able to distinguish the mirco-state of the actual configuration to a precision characterized by the volume of the patches. Going to this more fine grained description of the system corresponds to proceeding in the radial direction in the bulk. Finally, a completely fine grained (microscopic) description corresponds to the boundary. That is, the number of bulk inputs in each layer of the network counts the information about the system. This number increases in the radial direction and interpolates between the macroscopic and the microscopic description. 

In this picture, for a black hole in the center of AdS, coarse graining has to terminate, when the horizon of the black hole is reached. Therefore, not all patches can be joined and a nonvanishing (coarse grained) entropy emerges.

\section{Conclusions and Outlook}\label{Conclusions}

In this work, we introduced {\it classical holographic codes} and analyzed their properties. Interpreting the input of the codes as the bulk degrees of freedom and its output as the boundary degrees of freedom, a classical holographic code establishes a map between these. One of the main features of the codes is that a version of the Ryu-Takayanagi formula holds; the mutual information between a connected region $A$ on the boundary and its complement $A^c$ is given by the length of the minimal cut $\gamma_A$ that ends on the boundary of $A$. We defined the bulk region that is enclosed between $\gamma_A$ and the boundary region $A$ as the correlation wedge $C(A)$ of $A$.  We have shown that the bulk inputs contained in $C(A)$ can be reconstructed from the data in $A$. Furthermore, we have shown that a (bit flip) operation $O$, acting on any bulk input contained in $C(A)$, can be represented by multiple bit flips in the boundary region. We also established a notion of subregion duality. That is, we have shown that any operation $O$ acting on some input in the bulk can be represented in any boundary region $A$ that possesses a respective correlation wedge $C(A)$ such that the bulk input is contained in it. Finally, the additional bulk dimension can be interpreted as a coarse graining parameter that interpolates between the microscopic description at the boundary of AdS and the macroscopic description in the center of AdS.

We did not intend to construct a purely classical toy model for the AdS/CFT correspondence. However, interestingly, all the features we described above are to be expected from AdS/CFT. Furthermore, these are the features that are modeled by quantum error-correcting codes, such as the ones in \cite{Almheiri:2014lwa, Pastawski:2015qua}. Of course, there is the obvious caveat that the boundary theory is purely classical and by no means can approximate a quantum CFT. In particular, the entanglement structure of a quantum CFT is completely absent. Another shortcoming of the classical code is that bulk and boundary operations (bit flips) are rather simple compared to general operators appearing in a CFT. Finally, in our particular example, the center vertex has some shortcomings, as we described.  However, especially in the limit of large networks, the center vertex should not cause serious problems.

Even so there are these shortcomings in the construction, it is interesting to note that, by starting from a purely classical code, one can obtain all the AdS/CFT-like features, we outlined above. This shows that, given the geometric structure of the network, the scaling of the mutual information, {\it i.e.}, a version of the RT formula, and important bulk and operator reconstruction properties are due to the ``correlation structure'' and can exist even classically in the absence of quantum correlations, such as entanglement.

For the future, it would be interesting to generalize the bulk-to-boundary mappings of this work. In particular, it is an open question, whether suitable random networks could possess  properties similar  to the ones of  classical holographic codes. Recently, for random tensor networks, this was shown to be true \cite{Hayden:2016cfa}. Furthermore, it might be worthwhile to see whether classical analogs of the Witten-like diagrams introduced in \cite{Bhattacharyya:2016hbx} could be found for classical holographic codes. A further direction that might be worth following is to relate classical holographic codes to the reconstruction of Abelian subalgebras \cite{Harlow:2016vwg}. In this context, more elaborate codes might certainly be helpful. Another interesting project would be to find a connection between classical holographic codes and existing probabilistic codes used for error correction that, \textit{e.g.}, can be related to spin glass models \cite{sourlas_spin-glass_1989}.

\acknowledgments

We would like to thank Ilka Brunner, Dieter L\"ust, Yasser Omar, Erik Parr, and Cornelius Schmidt-Colinet for useful discussions and/or comments. Furthermore, we thank Robert Helling for a very useful comment. 

Both authors contributed equally to this work. The work of E.B. was supported by the Excellence Cluster Universe. The work of B.R. was supported by FCT through scholarship SFRH/BD/52651/2014. Furthermore, B.R. thanks the support from Funda\c{c}\~{a}o para a Ci\^{e}ncia e a Tecnologia, namely through programmes PTDC/POPH/POCH and projects UID/EEA/50008/2013, IT/QuSim, IT/QuNet, ProQuNet, partially funded by EU FEDER, from the EU FP7 project PAPETS (GA 323901), and from the JTF project NQN (ID 60478).

\appendix

\section{Some properties of $\mathfrak{S}$}\label{properties}

In this appendix, we discuss some of the properties of the set $\mathfrak{S}$ that we use in section \ref{fullnetwork}. First we consider bipartitions into substrings of length three. We require that the bipartitions are maximally correlated as quantified by the mutual information. The maximal possible value for the mutual information is, in this case, $3\log(2)$ which also is the maximal possible Shannon entropy of a string of three bits. Consider, for example,   the bipartition into input and output bits of the $3\rightarrow 3$ mapping, \textit{i.e.} into the strings $s_\text{in} = i_1i_2i_3$ and $s_\text{out} = i_4i_5i_6$, where $i_n$ is the $n$th bit in the full strings. We require $S_S(s_\text{in}) = 3 \log(2)$, such that 
\begin{equation}\label{condition}
 0 = S_S(s_\text{out}) - S_S(\mathfrak{S}).
\end{equation}
Since $S_S(s_\text{out})\le3\log(2)$ and $S_S(\mathfrak{S})\ge 3\log(2)$,  condition (\ref{condition}) can be satisfied only if $S_S(s_\text{out}) = 3\log(2) = S_S(\mathfrak{S})$. It follows that $|\mathfrak{S}| = N = e^{S_S(\mathfrak{S})} = 2^3 = 8$ and the $3\rightarrow3$ map is bijective. In general, it is true that
\begin{itemize}
\item[(I)] the knowledge of three neighboring edge bits gives full information about the three complementary bits.
\end{itemize}

Let us next consider bipartitions into a single bit and the remaining five bits. The maximal possible mutual information between these bipartitions is $\log(2)$. Since we already know that the Shannon entropy of the set $\mathfrak{S}$ is $3\log(2)$, we can conclude that the entropy of any single bit must be $\log(2)$ and that of any substring of five bits has to be $3\log(2)$. From the latter it follows that no two substrings of length five can be the same. One can also show that no two single bits $i_A$ and $i_B$ in $\mathfrak{S}$ can be correlated by deriving their mutual information 
\begin{equation}
\begin{split}
  I_\text{cl.}(i_A,i_B) &= S_S(i_A) + S_S(i_B) - S_S(i_A\cup i_B) \\&= \log(2)+\log(2)-2\log(2) = 0\,. 
\end{split}\label{eq:11mutinf}
\end{equation}
As a consequence of the fact that the  mutual information, $I_\text{cl.}(i_A,i_B)$, vanishes,
\begin{itemize}
\item[(II)] no information about any other single bit can be obtained by the knowledge of one particular bit. 
\end{itemize}

Finally, we consider the case of bipartitions into strings of length two and their complement. In this case, the maximal value for the mutual information is $2\log(2)$. As before, one can show that the Shannon entropy of two bits is always $2\log(2)$ (we already used this in \eqref{eq:11mutinf}) and the entropy of four bits has to equal $3\log(2)$, such that any two substrings of length four have to be different. The mutual information between two bits and a third bit vanishes if their union or their complement contains only neighboring bits. In particular, it follows that two neighboring edge bits never reveal information about bits next to them and, in general, two bits can at most give one other bit with certainty. 

A further consequence of demanding that any bipartition of $\mathfrak{S}$ into substrings of nonequal size is maximally correlated is that the tripartite information $I_3(A,B,C)$ that is defined as
\begin{equation}
I_3(A,B,C)=S_S(A)+S_S(B)+S_S(C)-S_S(AB)-S_S(AC)-S_S(BC)+S_S(ABC)\, ,
\end{equation}
where $A, B, C$ denote arbitrary subsets of neighboring  bits, is nonpositive, $I_3(A,B,C)\leq 0$. For all cases except for $|A|=|B|=|C|=2$, this can be shown using the upper bound $I_3(A,B,C)\leq \min \{I_\text{cl.}(A,B), I_\text{cl.}(B,C), I_\text{cl.}(A,C)\}$ that was obtained \textit{e.g.} in \cite{yeung1991new}. In the special case of a split in three sets of equal cardinality with $|A|=|B|=|C|=2$, it follows from the fact that $I_3(A,B,C)=I_\text{cl.}(A,B)+I_\text{cl.}(A,C)-S(A)=0$. The physical interpretation of $I_3(A,B,C)\leq 0$ is that the mutual information between any pair of subsets $A$, $B$, and $C$ increases once the other random variable is known.  Interestingly, it was shown that, for boundary regions $A$, $B$, and $C$, $I_3(A,B,C)\leq 0$ holds in AdS/CFT \cite{Hayden:2011ag}.

\section{Proof of a version of the RT formula for classical holographic codes}\label{proof}

In this appendix, we prove a version of the RT formula (\ref{classicalRTformula}) for classical holographic codes. Therefore, we first argue that the mutual information of a connected region $A$ and its complement is bounded from above by the length of the minimal cut $\gamma_A$, \textit{i.e.},
\begin{equation}\label{lower bound}
I_{cl}(A,A^c)\le |\gamma_A|\,. 
\end{equation} 
It is evident that all correlations in the system must be generated in the interior of the bulk and are transported by the network to the boundary. If we consider an arbitrary cut through the network whose ends coincide with the boundary of the interval, then all correlations between regions $A$ and $A^c$ are transmitted through the edges that are crossed by the cut. Of course, that is also true for the minimal cut $\gamma_A$ and, since every edge can at most transfer one bit of information, the amount of correlation (or shared information) is bounded from above by the length of this cut. Therefore, bound (\ref{lower bound}) holds.   
\begin{figure}[h]
 \centering
 \includegraphics[width=12cm]{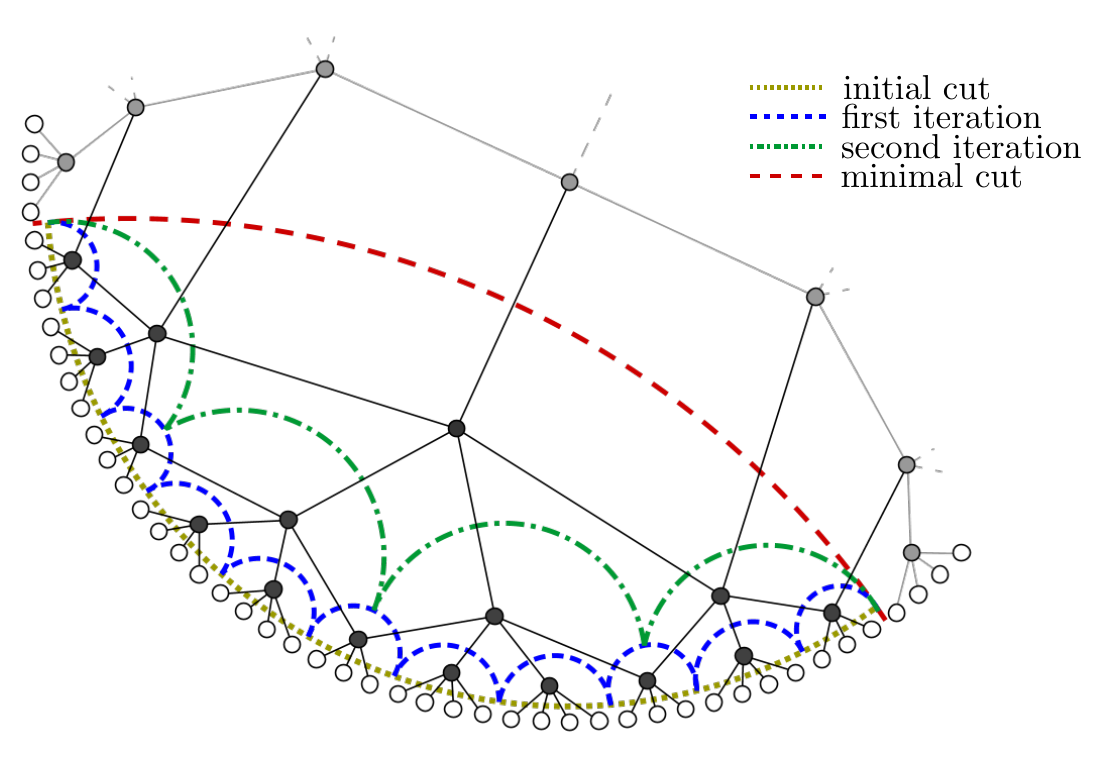}
 \caption{Visualization of the algorithm that constructs the minimal cut (red curve) for a boundary region. The algorithm starts from a cut that divides the bits in that boundary region from the remaining system (initial cut). Then, for each vertex, it evaluates how many edges belonging to the vertex are crossed by this cut. If this number is larger than or equal to three, the cut is moved across the vertex such that it cuts all edges of this vertex that have previously not been crossed (in the first iteration that results in the blue cut). Subsequently, it takes the new cut as starting point. The algorithm terminates, when the cut is minimal (red cut).}
 \label{fig:mincut}
\end{figure}

In the case of  classical holographic codes, the upper bound (\ref{lower bound}) for the mutual information is saturated, as we show next. The general idea of the proof is that any of the bits that are transferred through an edge crossed by the minimal cut $\gamma_A$ can be reconstructed with certainty from either side. Furthermore, there is no correlation between the edge bits crossed by $\gamma_A$. Then each of the bits has to carry one bit of shared information and hence contribute to the mutual information by one. In consequence, the mutual information is given by the length of the minimal cut $\gamma_A$ and the version of the RT formula \eqref{classicalRTformula} holds.

One can convince oneself that this statement is true by considering an algorithm for constructing the minimal cut that was also presented in \cite{Pastawski:2015qua}. Given some connected region of the boundary, the algorithm starts with the cut that crosses all open edges at the boundary. The algorithm then proceeds in the following way: It lets the cut jump over a vertex if at least three edges of one vertex are crossed by the cut. After the jump it crosses all the edges of the vertex that were not crossed before. Then, given the new cut, it starts again. The algorithm stops when the cut is minimal; cf.\ figure \ref{fig:mincut}. From that it is clear that each bit flowing through any edge crossed by a cut constructed in this way can be reconstructed from the bits of the boundary region it starts from. This directly follows by applying property (I) in every step of the algorithm. 

In most cases the minimal surface constructed from a connected region $A$ on the boundary and the one from its complement coincide. However, as also mentioned in \cite{Pastawski:2015qua}, there is the possibility that these do not coincide. If the minimal cut is unique, we certainly can construct its edge from both boundary regions.

Next, we argue that the edge bits that are crossed by a \textit{unique} minimal cut cannot be correlated. Therefore, we show that no information about an edge bit can be obtained by the knowledge of any subset of the remaining edge bits. 

\begin{figure}[ht]
 \centering
\includegraphics[width=.8\textwidth]{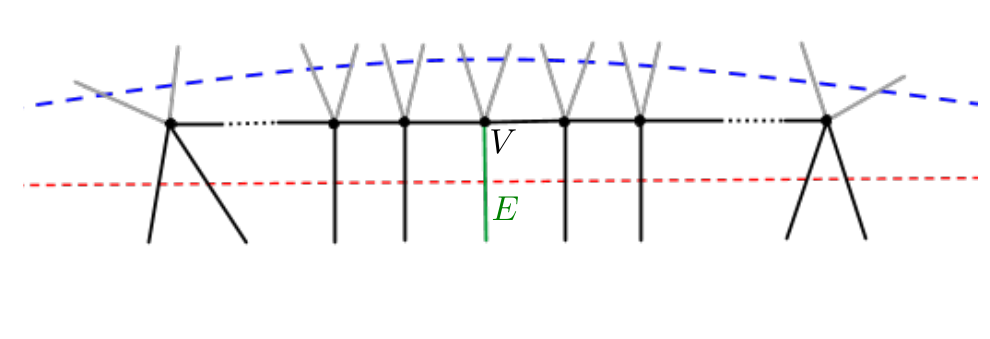}
 \caption{Illustration of the reasoning about the correlation of crossed edge bits. We assume the green edge bit through the red minimal cut can be constructed. The black edges are needed to gather information about the green one. The grey edges are pointing away from the minimal cut and can give no information. The blue cut is the minimal cut constructed -- using algorithm illustrated in figure \ref{fig:mincut} -- from the complementary boundary. The two minimal cuts do not coincide. Hence, if we assume some bits are correlated the minimal cut cannot be unique.}
 \label{fig:proof}
\end{figure}

First, let us assume the contrary, \textit{i.e.}, one can obtain information about a crossed edge bit $E$ from the knowledge of other crossed edge bits. In figure \ref{fig:proof}, which illustrates our proof, this is the green edge. Now fix the vertex $V$ from which one assumes to get information about $E$. Edges at that vertex that point ``deeper into the bulk'', and hence away from the minimal cut, cannot carry information about any other leg crossed by the minimal cut simply because their distance through the network to them is too large. In figure \ref{fig:proof}, these are the gray edges. Now there are two possibilities: either one of the two neighboring edges also crosses the minimal cut and the other edge goes parallel to it or both neighboring edges go parallel to it, where the latter is shown in figure \ref{fig:proof}. In both cases we need the knowledge of at least one edge going parallel to the minimal cut, because property (II) tells us that we need at least two bits to reconstruct a third one. Let us focus on this parallel edge and ask how to obtain the bit associated with it. Again because of property (II) we need to know at least two edge bits from the other vertex it is connected to, and again there are two possibilities: either there are two edges crossing the minimal cut or we have one edge crossing the minimal cut and another one going parallel to it. As before, we can get no information from edges pointing deeper into the bulk. We can conclude that one requires the knowledge from another parallel edge if there are not two known and necessarily neighboring edge bits crossed by the minimal cut.   

This logic stays true for any parallel edge bit and, hence, if we assume that we can reconstruct $E$ then there need to be two neighboring edge bits crossed by the minimal cut before we reach the boundary in both directions (which includes the possibility that $E$ itself is one of these bits). If not, then we would need a parallel edge at the boundary that by requirement is not known. This is also shown in figure \ref{fig:proof}: to construct $E$ one needs to know all the (black) crossed edge bits to construct the (black) parallel edge bits, where finally the two parallel edge bits next to $E$ are needed to construct $E$ itself. In summary, we need a ``chain'' of parallel edges, where the chain ends in both directions with two neighboring crossed edges.  

The crucial caveat is that the minimal cut \textit{cannot} be unique in the above situations! Simply consider the minimal cut whose construction started at the boundary in the direction of the parallel edges. This cut cannot jump over the vertices that are connected to the previously considered parallel edges, because there are always less than three edges pointing in the direction of the boundary. This is, in particular, the case at the two ends of the above ``chain''. This is also shown in figure \ref{fig:proof}, where the blue cut can only cross the gray legs. It cannot jump over the vertices to finally coincide with the red minimal cut. This now shows that the edges of a \textit{unique} minimal cut cannot be constructed from the knowledge of any subset of other crossed edges and, hence, none of them can be correlated. Together with the fact that each edge crossed by the minimal cut can be reconstructed from either side, this finishes the proof of the RT formula (\ref{classicalRTformula}).

\vspace{-.4cm}
\begin{flushright}
 $\square$
\end{flushright}

Note that there are still the cases left, where the minimal surface is not unique. From the argumentation above it becomes clear that for those the mutual information is smaller than the length of the minimal cut.

All these results are supported by numerical checks up to the fourth layer of the network.\footnote{The numerical results also suggest that in the case of two different minimal cuts coming from $A$ and $A^c$, the mutual information is given by the length of the smaller cut minus the number of connected regions between the two cuts.}

\bibliographystyle{JHEP}
\providecommand{\href}[2]{#2}\begingroup\raggedright

\end{document}